\documentclass[12pt]{article}             
\begin{document}
\begin{center}
{\bf Minkovskii-type inequality for arbitrary density matrix of composite and noncomposite systems}\\
V. N. Chernega, O.~V.~Man'ko, V. I. Man'ko\\
 P.N.~Lebedev Physical Institute, Russian Academy of Sciences\\
      Leninskii Prospect, 53, Moscow 119991, Russia \\
      Email: omanko@sci.lebedev.ru
\end{center}
\begin{abstract}
New kind of matrix inequality known for bipartite system density matrix is obtained for arbitrary density matrix of composite or noncomposite qudit systems including the single qudit state. The examples of two qubit system and qudit with $j=3/2$ are discussed.

\end{abstract}
\vskip1cm

 \vspace{0.4cm}

\section{Introduction}
The quantum correlations for multipartite qudit systems are partially characterized by some inequalities for density matrices of the systems. For example, the system of two qubits has the density matrix which in case of separable state of the system obeys the Bell inequality \cite{Bell1964,CHCH}. The violation of the inequality provides the characteristics of the entanglement in the two qubit system which is related to the degree of correlations between qubits and correlations can be associated with value of Cirelson bound \cite{Cir}. On the other hand there exist the entropic and information inequalities, e.g. subadditivity condition which is inequality for von Neumann entropies of the bipartite system and its two subsystem states \cite{fromRusk}. For three-partite systems there exists strong subadditivity condition which is the inequality for the von Neumann entropies of the composite system and its subsystems \cite{LiebRuskai,Ruskai}. The subadditivity condition are also valid for Shannon entropies \cite{Shannon} of the bipartite and three-partite systems, respectively. The nonnegativity of the Shannon mutual information and  quantum mutual information and conditional classical and quantum mutual informations follow from the subadditivity and strong subadditivity conditions. Recently the portrait qubit and qudit map was introduced to study the entanglement phenomenon \cite{Vovf,Lupo}. This method is appropriate to study quantum correlations in framework of tomographic probability representation of quantum states \cite{Mancini96,ovmvimjrlr1997,IbortPhys2007,NuovoCim,maps}. In this representation which is valid for both discrete and continious variables \cite{Mancini96} the spin states (qudit states) are identified with fair tomographic probability distributions \cite{DodPLA,OlgaJETP,OlgaBregence,FilipJRLR}. In view of this the standard formulas for classical probability distributions like entropies can be easily compared with corresponding quantum ones \cite{FoundFi,MamaMendesJRLR}. Using the approach based on the portrait method which in fact is the positive map approach it was shown \cite{Vovf,Lupo,OlgaVova,OlgaVova1,RitaPS2014,OlgaVovaVI} that the entropic inequalities valid for composite systems can be extended to the arbitrary systems including systems without subsystems. In \cite{CorlenLieb2007,20pr,2014} some inequalities associated with positive operators acting in the Hilbert space which has the structure of tensor product of Hilbert spaces were studied. The aim of our work is to obtain new matrix inequalities for density matrices of the qudit states of both composite and noncomposite quantum systems which do not depend on the tensor product structure of Hilbert space. We follow the approach of \cite{Vovf,Lupo,OlgaVova,OlgaVova1,RitaPS2014,OlgaVovaVI} based on the portrait positive map method.

The paper is organized as follows. In Sec. 2 we formulate new inequality for the Hermitian matrix which corresponds to operator inequality for bipartite quantum system given in \cite{CorlenLieb2007}. In Sec. 3 we consider example of this inequality for $4\times4$ Hermitian matrix and for density matrix of two-qubit state and the single qudit with $j=3/2$. The conclusion and perspectives are given in Sec. 4.

\section{Inequality for $N$ $\times$ $N$ - Hermitian matrix}
We present now the new inequality for density $N$ $\times$ $N$ matrix, i.e. $\rho^+=\rho,\, \mbox{ Tr}\rho=1,\,\rho\geq0$. Let $N$=$n$$m$ where $n$ and $m$ are integers. Let us present the matrix $\rho$ in block form
\begin{equation}\label{LM1}
\rho=\left(\begin{array}{cccc}
a_{11}&a_{12}&\cdots&a_{1n}\\
a_{21}&a_{22}&\cdots&a_{2n}\\
\cdots &\cdots &\cdots&\cdots\\
a_{n1}&a_{n2}&\cdots&a_{nn}\end{array}\right).
\end{equation}
Here the blocks $a_{jk}$ $(j,k=1,2,\ldots,n)$ are the
$m$$\times$$m$ matrices. For arbitrary real number $p$ we introduce the matrix $\rho^p$ and present it in block form \begin{equation}\label{LM2}
\rho^p=\left(\begin{array}{cccc}
a_{11}(p)&a_{12}(p)&\cdots&a_{1n}(p)\\
a_{21}(p)&a_{22}(p)&\cdots&a_{2n}(p)\\
\cdots &\cdots &\cdots&\cdots\\
a_{n1}(p)&a_{n2}(p)&\cdots&a_{nn}(p)\end{array}\right).
\end{equation}
The blocks $a_{jk}(p)$ are the $m$ $\times$ $m$ matrices which depend on the parameter $p$. New inequality which is valid for arbitrary $N$ $\times$ $N$ - matrix $\rho$ reads
\begin{equation}\label{LM3}
\left[\mbox{Tr}\left(\sum_{j=1}^n a_{jj}\right)^p\right]^{1/p}\leq\mbox{Tr}\left[\left(\begin{array}{cccc}
\mbox{Tr}a_{11}(p)&\mbox{Tr}a_{12}(p)&\cdots&\mbox{Tr}a_{1n}(p)\\
\mbox{Tr}a_{21}(p)&\mbox{Tr}a_{22}(p)&\cdots&\mbox{Tr}a_{2n}(p)\\
\cdots &\cdots &\cdots&\cdots\\
\mbox{Tr}a_{n1}(p)&\mbox{Tr}a_{n2}(p)&\cdots&\mbox{Tr}a_{nn}(p)\end{array}\right)^{1/p}\right].
\end{equation}
This inequality is valid for $p\geq 1$. In this inequality the $n\times n$-matrix in right-hand side has matrix elements $\mbox{Tr}a_{jk}(p)$. If $0\leq p\leq1$ the inequality (\ref{LM3}) reverses.

If $N\neq n m$ we use the integer $N'=N+s$ such that $N'=n m$ and consider the density $N'\times N'$ matrix $\rho'$ of the form
\begin{equation}\label{LM4}
\rho'=\left(\begin{array}{cccc}\rho&0\\0&0\end{array}\right)=\left(\begin{array}{cccc}
a_{11}&a_{12}&\cdots&a_{1n}\\
a_{21}&a_{22}&\cdots&a_{2n}\\
\cdots &\cdots &\cdots&\cdots\\
a_{n1}&a_{n2}&\cdots&a_{nn}\end{array}\right).
\end{equation}
Here $a_{jk}$ are blocks which provide the representation of density $N'\times N'$ matrix $\rho'$. Then the inequality (\ref{LM3}) takes place for the blocks associated with the matrix $\rho'$ by (\ref{LM4}). The new inequality (\ref{LM3}) is obtained on the base of inequality obtained in \cite{CorlenLieb2007,20pr,2014} for the density operator of a bipartite quantum system. But the new inequality (\ref{LM3}) is valid for arbitrary density matrices of multipartite qudit systems including the single qudit density matrix. If the density matrix $\rho$ is diagonal matrix the inequality (\ref{LM3}) provide the inequalities for arbitrary probability vectors.

In fact, let us denote the diagonal elements of the matrix $\rho$ as
\[(a_{jj})_{\alpha}=P_{j\alpha}, \quad \alpha=1,2,\ldots m.\]
The nonnegative numbers $P_{11},P_{12},\ldots,P_{1m},P_{21},P_{22},\ldots,P_{2m},\ldots,P_{n1},P_{n2},\ldots,P_{n m}$ can be considered as components of a probability $N$-vector $\vec P$. The inequality (\ref{LM3}) written in terms of the probability vector reads
\begin{equation}\label{LM5}
\left[\sum_{\alpha=1}^m\left[\left(\sum_{j=1}^n P_{j\alpha}\right)^p\right]\right]^{1/p}\leq\sum_{j=1}^n\left[\sum_{\alpha =1}^m  \left(P_{j\alpha}\right)^p\right]^{1/p},\quad p\geq 1.
\end{equation}
The reverse inequality holds for $0<p\leq 1$. If $N\neq n m$ we use $N'=N+s=n m$. The inequality (\ref{LM5}) for the probability $N'$-vector holds and the last $s$ components of the probability $N'$-vector are equal to zero. In fact we have the inequality (\ref{LM5}) for arbitrary number of nonnegative numbers which are not necessarily associated with a probability distribution. It is worthy to point out that the inequality (\ref{LM3}) we obtained for density $N\times N$ matrix $\rho$ is valid for any density matrix obtained from this one by all the permutations of indices $1,2,\ldots,N\rightarrow1_p,2_p,\ldots,N_p$. The same statement is true for the inequality (\ref{LM5}). More generally, the density matrix $\Phi(\rho)$ obtained from the initial matrix $\rho$ by means of arbitrary positive map $\rho\rightarrow\Phi(\rho)$ satisfies the inequality (\ref{LM3}). It is clear that one can use different decompositions of the integer $N=n m=n'm'$. It means that there exist different inequalities for the same density matrix $\rho$ corresponding to different product form of the numbers $N$ and $N'$.

For $N=n m$ we can extend the inequality (\ref{LM3}) to the case of arbitrary Hermitian $N\times N$ matrix $A$. Let $A=A^\dag$ and the matrix $A$ has the block form corresponding to decomposition $N=n m$
\begin{equation}\label{LM6}
A=\left(\begin{array}{cccc}
a_{11}&a_{12}&\cdots&a_{1n}\\
a_{21}&a_{22}&\cdots&a_{2n}\\
\cdots &\cdots &\cdots&\cdots\\
a_{n1}&a_{n2}&\cdots&a_{nn}\end{array}\right),
\end{equation}
Let $x_0$ be the minimal eigenvalue of the matrix $A$. For arbitrary $x\geq x_0$ such that $x+x_0\geq0$ we introduce the nonnegative Hermitian matrix
\begin{equation}\label{LMa}
A(x)=A+x1_N.
\end{equation}
The matrix $A(x)$ has the block form
\begin{equation}\label{LM7}
A(x)=\left(\begin{array}{cccc}
a_{11}+x1_m&a_{12}&\cdots&a_{1n}\\
a_{21}&a_{22}+x1_m&\cdots&a_{2n}\\
\cdots &\cdots &\cdots&\cdots\\
a_{n1}&a_{n2}&\cdots&a_{nn}+x1_m\end{array}\right).
\end{equation}
Then the matrix $(A(x))^p$ can be presented in block form
\begin{equation}\label{LM8}
(A(x))^p==\left(\begin{array}{cccc}
a_{11}(x,p)&a_{12}(x,p)&\cdots&a_{1n}(x,p)\\
a_{21}(x,p)&a_{22}(x,p)&\cdots&a_{2n}(x,p)\\
\cdots &\cdots &\cdots&\cdots\\
a_{n1}(x,p)&a_{n2}(x,p)&\cdots&a_{nn}(x,p)\end{array}\right).
\end{equation}
For $N=n m$ the new inequality which holds for arbitrary Hermitian $N\times N$ matrix $A$ reads
\begin{equation}\label{LM9}
\left[\mbox{Tr}\left(\sum_{j=1}^n a_{jj}(x)\right)^p\right]^{1/p}\leq\mbox{Tr}\left[\left(\begin{array}{cccc}
\mbox{Tr}a_{11}(x,p)&\mbox{Tr}a_{12}(x,p)&\cdots&\mbox{Tr}a_{1n}(x,p)\\
\mbox{Tr}a_{21}(x,p)&\mbox{Tr}a_{22}(x,p)&\cdots&\mbox{Tr}a_{2n}(x,p)\\
\cdots &\cdots &\cdots&\cdots\\
\mbox{Tr}a_{n1}(x,p)&\mbox{Tr}a_{n2}(x,p)&\cdots&\mbox{Tr}a_{nn}(x,p)\end{array}\right)^{1/p}\right].
\end{equation}
This inequality holds for $p\geq1$. The inequality reverses for $0\leq p\leq1$. In case $N\neq n m$ we use $N'=N+s=n m$ and the new inequality for the Hermitian matrix
\begin{equation}\label{LM10}
A'=\left(\begin{array}{cc}
A&0\\
0&0\end{array}\right)
\end{equation}
presented in the block form (\ref{LM6}) can be given by the above inequality (\ref{LM9}). Using the diagonal Hermitian matrix $A$ one can write down inequality for arbitrary finite set of $N=n m$ real numbers $P_{11},P_{12},\ldots, P_{1m},P_{21},P_{22},\ldots,P_{2m},\ldots, P_{n1},P_{n2},\ldots, P_{n m}$. It has the form of inequality for two functions ${\cal P}_1(x,p)$ and ${\cal P}_2(x,p)$, i.e.
\begin{eqnarray}
&&{\cal P}_1(x,p)\leq {\cal P}_2(x,p), \quad p\geq1, \nonumber\\
&&{\cal P}_1(x,p)\geq {\cal P}_2(x,p), \quad 0<p\leq1.\label{LM11}
\end{eqnarray}
In this inequality
\begin{eqnarray}
&&{\cal P}_1(x,p)=\left\{\sum_{\alpha=1}^m\left[\left(n x+\sum_{j=1}^n P_{j\alpha}\right)^p\right]\right\}^{1/p}, \nonumber\\
&&{\cal P}_2(x,p)=\sum_{j=1}^n\left\{\left[\left(\sum_{\alpha=1}^m P_{j\alpha}\right)+m x\right]^p\right\}^{1/p}\label{LM11a}
\end{eqnarray}
For reals such that $P_{j\alpha}\geq0$ and $\sum_{j=1}^n\sum_{\alpha=1}^m P_{j\alpha}=1$ the inequality (\ref{LM11}) can be interpreted as the inequality for probability vector which holds for arbitrary $x\geq0$.

Some information on the correlations in the system of qudits including the case of single qudit is available in the  difference of terms in inequality (\ref{LM3})
\begin{eqnarray}\label{LM12}
{\cal J}(p)=\mbox{Tr}\left[\left(\begin{array}{cccc}
\mbox{Tr}a_{11}(p)&\mbox{Tr}a_{12}(p)&\cdots&\mbox{Tr}a_{1n}(p)\\
\mbox{Tr}a_{21}(p)&\mbox{Tr}a_{22}(p)&\cdots&\mbox{Tr}a_{2n}(p)\\
\cdots &\cdots &\cdots&\cdots\\
\mbox{Tr}a_{n1}(p)&\mbox{Tr}a_{n2}(p)&\cdots&\mbox{Tr}a_{nn}(p)\end{array}\right)^{1/p}\right]-\left[\mbox{Tr}\left(\sum_{j=1}^n a_{jj}\right)^p\right]^{1/p}\geq0,\quad p\geq1.
\end{eqnarray}
For bipartite system the function ${\cal J}(p)$ is additional characteristics of correlations to the mutual  information given by the subadditivity condition terms.

\section{The inequalities for Hermitian $4\times4$ matrices}
Let us illustrate the inequalities on example of $4\times4$-matrices. In this case the $4\times4$-matrix $A$ has the $2\times2$ blocks
\begin{equation}\label{EX1}
a_{11}=
\left(\begin{array}{cc}
\rho_{11}&\rho_{12}\\
\rho_{21}&\rho_{22}\end{array}\right),\quad
a_{12}= \left(\begin{array}{cc}
\rho_{13}&\rho_{14}\\
\rho_{23}&\rho_{24}\end{array}\right),\quad
a_{21}=
\left(\begin{array}{cc}
\rho_{31}&\rho_{32}\\
\rho_{41}&\rho_{42}\end{array}\right),\quad
a_{22}=
\left(\begin{array}{cc}
\rho_{33}&\rho_{34}\\
\rho_{43}&\rho_{44}\end{array}\right).\end{equation}
The matrix $A(x)$ reads
\begin{equation}\label{EX2}
A(x)=\left(\begin{array}{cccc}
\rho_{11}+x&\rho_{12}&\rho_{13}&\rho_{14}\\
\rho_{21}&\rho_{22}+x&\rho_{23}&\rho_{24}\\
\rho_{31}&\rho_{32}&\rho_{33}+x&\rho_{34}\\
\rho_{41}&\rho_{42}&\rho_{43}&\rho_{44}+x
\end{array}\right).
\end{equation}
The matrix $(A(x))^p$ has the block form
\begin{equation}\label{EX3}
(A(x))^p=\left(\begin{array}{cc}
a_{11}(x,p)&a_{12}(x,p)\\
a_{21}(x,p)&a_{22}(x,p)
\end{array}\right).
\end{equation}
The inequality (\ref{LM9}) has the form
\begin{equation}\label{EX16}
\left[\mbox{Tr}\left(\begin{array}{cc}
\rho_{11}+\rho_{33}+2x&\rho_{12}+\rho_{34}\\
\rho_{21}+\rho_{43}&\rho_{22}+\rho_{44}+2x\\
\end{array}\right)^p\right]^{1/p}\leq\mbox{Tr}\left[\left(\begin{array}{cc}
\mbox{Tr}a_{11}(x,p)&\mbox{Tr}a_{12}(x,p)\\
\mbox{Tr}a_{21}(x,p)&\mbox{Tr}a_{22}(x,p)\\
\end{array}\right)^{1/p}\right], \quad p\geq1.
\end{equation}
If the Hermitian matrix $A$ is nonnegative and $\mbox{Tr}A=1$ it can be interpreted as a density matrix either of two-qubit state or the state of qudit with $j=3/2$.

For diagonal density $4\times4$ matrix with eigenvalues $p_{11},\,p_{12},\,p_{21},\,p_{22}$ the inequality reads ($x=0$)
\begin{equation}\label{EX17}
\left[\left(p_{11}+p_{21}\right)^p+\left(p_{12}+p_{22}\right)^p\right]^{1/p}\leq \left(p_{11}^p+p_{12}^p\right)^{1/p}+\left(p_{21}^p+p_{22}^p\right)^{1/p}, \quad p\geq1.
\end{equation}
One can check that for $p=2$ the above inequality is equivalent to inequality $a^2+b^2\geq2a b$. The function ${\cal J}(p)$ for this case reads
\begin{equation}\label{EX18}
{\cal J}(p)=\left(p_{14}^p+p_{12}^p\right)^{1/p}+\left(p_{21}^p+p_{22}^p\right)^{1/p}
-\left[\left(p_{11}+p_{21}\right)^2+\left(p_{12}+p_{22}\right)^p\right]^{1/p}\geq0, \quad p\geq1.
\end{equation}
The mutual information for this case has the form
\begin{eqnarray}
&&I=p_{11}\ln p_{11}+p_{12}\ln p_{12}+p_{21}\ln p_{21}+p_{22}\ln p_{22}-\left(p_{11}+p_{12}\right)\ln\left(p_{11}+p_{12}\right)-\left(p_{21}+p_{22}\right)\ln\left(p_{21}+p_{22}\right)\nonumber\\
&&- \left(p_{11}+p_{21}\right)\ln\left(p_{11}+p_{21}\right)-\left(p_{12}+p_{22}\right)\ln\left(p_{12}+p_{22}\right)\geq0\label{EX19}
\end{eqnarray}
The inequalities (\ref{EX18}) and (\ref{EX19}) are compatible.

\section*{Conclusion}
To conclude we point out the main results of our work. We obtained for arbitrary system of qudits, including single qudit case, the inequalities for the system-state density matrix which is equivalent to known inequalities in case of bipartite quantum system. We obtained new simple inequality for arbitrary Hermitian $N\times N$-matrix. The inequality can be used to study the ground state energy property for the Hermitian Hamiltonian and the compatibility of this inequality with entropic and information inequalities which can be obtained for the Hamiltonian. These problems will be discussed in a future publication.

\end{document}